\documentclass[journal]{IEEEtran}

\usepackage{color,array,amsthm}
\usepackage{graphicx}
\usepackage{cite}
\usepackage[utf8]{inputenc}
 \usepackage{setspace}
\usepackage{amsmath,amssymb,amsfonts}
\usepackage{dirtytalk}
\usepackage{algorithmic}
\usepackage{graphicx}

\usepackage{textcomp}
\usepackage{xcolor}
\usepackage{amsfonts}
\usepackage{array,multirow,graphicx}

\usepackage{tikz}
\usetikzlibrary{tikzmark,decorations.pathreplacing,calc}
\DeclareSymbolFont{rmlargesymbols}{OMX}{mdbch}{m}{n}
\DeclareMathSymbol{\rmintop}{\mathop}{rmlargesymbols}{82}

\usepackage{siunitx}
\usepackage{gensymb}
\usepackage[english]{babel} % To obtain English text with the blindtext package
\addto\captionsenglish{\renewcommand*{\tablename}{TABLE}}

\usepackage{blindtext}
\newcommand{\RNum}[1]{\uppercase\expandafter{\romannumeral #1\relax}}

\def\BibTeX{{\rm B\kern-.05em{\sc i\kern-.025em b}\kern-.08em
    T\kern-.1667em\lower.7ex\hbox{E}\kern-.125emX}}
\usepackage{atbegshi}% http://ctan.org/pkg/atbegshi
\AtBeginDocument{\AtBeginShipoutNext{\AtBeginShipoutDiscard}}

\makeatletter
\renewcommand{\fnum@figure}{Fig. \thefigure}
\makeatother

\begin{document}

\title{On the Role of Reflectarrays for Interplanetary Links}

{\author{Eray~Guven,~\IEEEmembership{Member,~IEEE,} Pablo Camacho,~\IEEEmembership{Member,~IEEE,}\\ Elham Baladi,~\IEEEmembership{Member,~IEEE,} Gunes~Karabulut Kurt,~\IEEEmembership{Senior Member,~IEEE} 
}}

\thanks{This work was supported in part by the Canada Research Chair program. (Corresponding Author: {\it Eray Guven})

Eray Guven, Pablo Camacho, Elham Baladi and Gunes Karabulut Kurt are with the Department of Electrical Engineering and Poly-Grames Research Center,
Polytechnique Montréal, Montréal, QC, H3T 1J4, Canada, (e-mail: {guven.eray@polymtl.ca}, {pablo-emilio.camacho-prieto@polymtl.ca}, {elham.baladi@polymtl.ca}, {gunes.kurt@polymtl.ca}).}

%% \author{FOURTH D. AUTHOR}
%% \affil{University of Colorado, Colorado, USA}

%\receiveddate{Manuscript received XXXXX 00, 0000\\
%``This work was supported in part by the U.S. Department of Commerce under Grant BS123456.'' }
%% \accepteddate{XXXXX XX XXXX}
%% \publisheddate{XXXXX XX XXXX}

\maketitle

\setcounter{page}{1}

\begin{abstract}
Interplanetary links (IPL) serve as crucial enablers for space exploration, facilitating secure and adaptable space missions. An integrated IPL with inter-satellite communication (IP-ISL) establishes a unified deep space network, expanding coverage and reducing atmospheric losses. The challenges, including irregularities in charged density, hardware impairments, and hidden celestial body brightness are analyzed with a reflectarray-based IP-ISL between Earth and Moon orbiters. It is observed that $10^{-8}$ order severe hardware impairments with intense solar plasma density drops an ideal system's spectral efficiency (SE) from $\sim\!38~\textrm{(bit/s)/Hz}$ down to $0~\textrm{(bit/s)/Hz}$. An ideal full angle of arrival fluctuation recovery with full steering range achieves $\sim\!20~\textrm{(bit/s)/Hz}$ gain and a limited beamsteering with a numerical reflectarray design achieves at least $\sim\!1~\textrm{(bit/s)/Hz}$ gain in severe hardware impairment cases. 
\end{abstract}


\section{Introduction}
Deep space networks are becoming popular for exciting Earth-to-Mars \cite{abdelsadek2022future} and lunar communication \cite{balossino2022moon,plan2020nasa} with interplanetary links (IPL). Currently, the backbone ground-to-air (G2A) networks link a ground station on Earth to an orbiter around the target planet. This approach has multiple advantages, however a G2A network constraints the communication in several aspects. One major constraint is the coverage, which leaves out-of-coverage dark zones on the target planet (e.g., the far side of the Moon \cite{donmez2024continuous,israel2023lunanet}). A second constraint is the exposure to severe atmospheric conditions on Earth and space weather. For example, it is well known that in both terahertz (THz) and microwave spectra, there are windows of band gaps that are not reasonable to operate within the earth's atmosphere due to the absorption by atmospheric gases and scattering by the atmospheric particles.

\textcolor{black}{The strategic advantages of air-to-air communication are discussed in \cite{saeed2021point}, highlighting that the potential benefits are primarily focused on earth-based communication. Nevertheless,} an integrated IPL with an inter-satellite link (IP-ISL) can set a highly efficient and feasible deep-space network. One recent study proposes a relayed IP-ISL for increasing the coverage \cite{chen2019lunar}, and another study performs space debris exploration by using IP-ISL \cite{urieta2023intelligent}. A recent survey shows that IP-ISL can be the new backbone of the large-scale unified space communication networks \cite{alhilal2021roadmap}. In this regard, we evaluate the performance of an IP-ISL with zero misalignment and full controllability under hardware impairments, hidden celestial body noises, and a deep space channel and examine the significance of each factor on the spectral efficiency (SE) in perpetual Earth-to-Moon connectivity.

Hidden planets and celestial bodies are part of deep space and are yet to be explored. By October 2021, $19$ possible hidden planets were discovered due to the interaction with LOFAR \cite{callingham2021population}. So far, recent studies show that even the Solar system possibly has undiscovered cloud planets within the Oort cloud that we are unable to detect. Therefore, radar and radio telescopes still have limited capability for space surveillance.  In long-distance IP links, the presence of asteroids, meteors, and meteoroids exceeds expectations, and these numerous and varying-sized celestial bodies pose a potential threat to the communication link. Other than observable kilometer-sized objects such as space rocks, one study \cite{ito2010asymmetric} highlights the source of lunar craters by the near-earth asteroids, whereas another technical report \cite{suggs2015results} focuses on the impact of the average $260$ meteoroids with a mass greater than $1$ kg on the Moon per year. For such a long-distance communication between Earth and Moon orbits ($\sim\!384,400 $ km), randomly spread celestial bodies are detrimental noise sources for IP-ISL due to the application of high-gain antennas.

Under solar activities, the Earth-to-Moon communication channel is solely formed by the solar plasma. Due to the plasma density in free space, the refractions on the propagation path are exposed to charged particles. The plasma irregularities in near-earth aerial channels and satellites have been discussed in \cite{zakharenkova2016gps, sato2018imaging,nishimura2021cross}. Similarly, the phase deviation on ISL for optical ranging is shown in \cite{lu2021effects}. A directly related study analyzes the phase scintillations as Doppler noise and investigates the spectral broadening \cite{morabito2009spectral}. For this reason, we highlight the need for beam management design during the pointing, acquisition, and tracking (PAT) to suppress the impact of deep space channels and noise on astronomical distance \cite{hill2007autonomous}. The major problems and our solution approaches are given in Table I. 

Reflectarray antennas offer a simplified alternative to phased array antennas, known for their complexity and high associated costs at frequencies above microwave. This streamlined approach helps to maintain lower costs for unified space communication networks \cite{theoharis2022wideband}.
%The major problems and our solution approaches are given in Table  Aligning with this, a high gain phased array antenna is required for IP-ISL   
\begin{table}[!tb]
  \centering
  \caption{IP-ISL Model} 
  \setlength{\tabcolsep}{8pt}
  \resizebox{0.8\linewidth}{!}{
  \begin{tabular}{|c|l|}
    \hline
    \multicolumn{1}{|c|}{\textbf{Solution}} & \multicolumn{1}{c|}{Problem} \\ \hline
     \multirow{3}{*}{\rotatebox[origin=c]{0}{Thermal noise modeling}}
     & \textbullet\ Non-ideal hardware noises                      \\
     & \textbullet\ Hidden celestial bodies                      \\
     & \textbullet\ EM disturbances                      \\ \hline
     \multirow{3}{*}{\rotatebox[origin=c]{0}{Beam management}}
     & \textbullet\ AoA fluctuations                      \\
     & \textbullet\ Free space diffractions                      \\
     & \textbullet\ Antenna misalignments                      \\ \hline
  \end{tabular}
  }
\end{table}
A reflectarray is able to steer the beam on the broadside in spatial tracking by manipulating the phase distributions over the aperture cells. The planar surface is mounted on a dielectric substrate which provides mechanical support and \textcolor{black}{facilitates the mechanical fabrication.} The aperture is illuminated by an exterior feed antenna. \textcolor{black}{By leveraging electromagnetic waves in the near-field on a transceiver aperture with reflecting unit cells, a reflectarray can achieve high-precision beam management, similar to the use of reflecting intelligent surfaces (RIS). A recent study \cite{gros2021reconfigurable} presents an analytical model for millimeter-wave reflectarray configurations utilizing binary phase RIS.} Recently, the use of reflectarray for air-to-ground (A2G) satellite communication is justified by the flexibility, weight, and durability as discussed in \cite{baladi2021dual, baladi2022low} and \cite{chung2023antennas}. \textcolor{black}{In examining the impact of hardware impairments on metasurfaces, a recent study \cite{boulogeorgos2020much} shows that transceiver hardware imperfections create bottlenecks in RIS-assisted communications.}

\textcolor{black}{In the literature, a stochastic free space optics-based relayed system performance is analyzed under solar corona fading channel \cite{xu2023relay}, yet lower operation bands are neglected. The authors in \cite{guven2023multi} takes the lower operation bands into account to present an ISL design, but for only a single constellation. Similarly, the deep space channel is mostly considered as solely free space \cite{fang2023sensing}, which is not a realistic assumption for IPL. With the recent advancements in satellite compatible reflectarrays in X-band and THz \cite{su2019x} - \cite{da2018design}, we show the necessity of a precise beam management, even in the perfectly aligned and localized IP-ISL communication scenarios.}

\textcolor{black}{
A perpetual IP-ISL necessitates the provision of efficient spectrum use within constrained resource parameters. Thus, a realistic signal-to-noise ratio (SNR) maximization on IP-ISL is only possible with beam management and thermal noise modeling. Accordingly, we first model the noise spectral density for end-to-end IP-ISL under weak \& strong electromagnetic (EM) disturbances, and then we find the fluctuations on angle-of-arrival (AoA) to be able to control the incident wave towards feed on reflectarrays.}

The contributions of this study are the followings:
\begin{itemize}
    \item An IP-ISL between Earth-to-Moon orbiters is constructed on a deep space channel under the presence of both known and unknown celestial bodies, hardware impairments, and dynamic free-space plasma density with ideal PAT with zero misalignment and full controllability.

    \item A feed-offset planar reflectarray engineered for low payload and high-gain is utilized for the IP-ISL. SE gain of the reflectarray is shown for both ideal and non-ideal cases.  

    \item The impact of hidden celestial bodies over IP-ISL is analyzed as a source of thermal noise on the main beamwidth of the designed antenna.  
        
    \item The variance of the AoA fluctuations caused by the solar activities are derived in reflectarray use. 

    \item The advantages and disadvantages of utilizing the X-band in IP-ISL are examined. A comparative analysis is conducted, considering external thermal noise and system hardware impairment, to assess their impact on SE. 
    
\end{itemize}

\textcolor{black}{The direct application of reflectarray technology paves the path for large-scale IP-ISL deployments at a reduced cost and with low size, weight, and power (SWaP), simplifying beam management through passive reflecting unit cells. This stands in contrast to phased arrays, which require active phase shifters and amplifiers.}

Accordingly, an Earth-to-Moon IP-ISL thermal noise and channel model is described in Section \ref{sec2}. The utilization of a lossless reflectarray design for beamsteering applications is discussed in Section \ref{sec3}, and numerical analysis methodologies and simulations are given in Section \ref{sec4}. We conclude our findings in Section \ref{sec5}.

\section{System Description}
\label{sec2}
\begin{figure}[]
    \centering
    \includegraphics[width=0.8\linewidth]{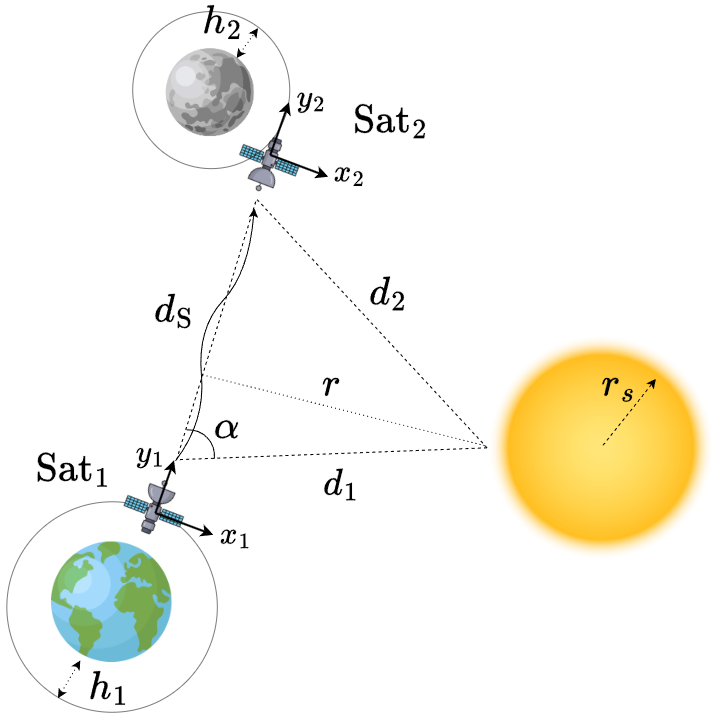}
    \caption{An IP-ISL of $\textrm{Sat}_1-\textrm{Sat}_2$ on free space including the Sun, Earth and the Moon. Assuming orbital inclinations to be zero allows a simplified 2D model in the $xy-$ plane. The diffracted trajectory along the $\textrm{Sat}_1-\textrm{Sat}_2$ path due to $\Delta \epsilon$, leading to $\delta\theta$ fluctuation on AoA.}
    %\vspace{-5mm}
    \label{ISLGeo}
\end{figure}
A transmitter satellite $\text{Sat}_1$ orbiting around the Earth and a receiver satellite $\text{Sat}_2$ orbiting around the Moon for IP-ISL during the coverage timeline are shown in Figure \ref{ISLGeo}. The orbiters are positioned at heights $h_1$ and $h_2$, near geosynchronous orbit around their respective home planets, thereby minimizing atmospheric effects like ionization, molecular absorption, and attenuation. Both satellites assume knowledge of the orbiter positions $p_{S_1}$ and $p_{S_2}$, ensuring beam alignment is unaffected by mobility. First, the IP-ISL is exposed to hardware impairments, thermal noises caused by the celestial body brightness, and the cosmic microwave background. Additionally, the IP-ISL channel is determined by the sun plasma density constituted by the solar corona \cite{muhleman1977electron}. Correspondingly, the noise and channel models are given in the following sections.
\begin{figure}[]
    \centering    \includegraphics[width=0.5\textwidth]{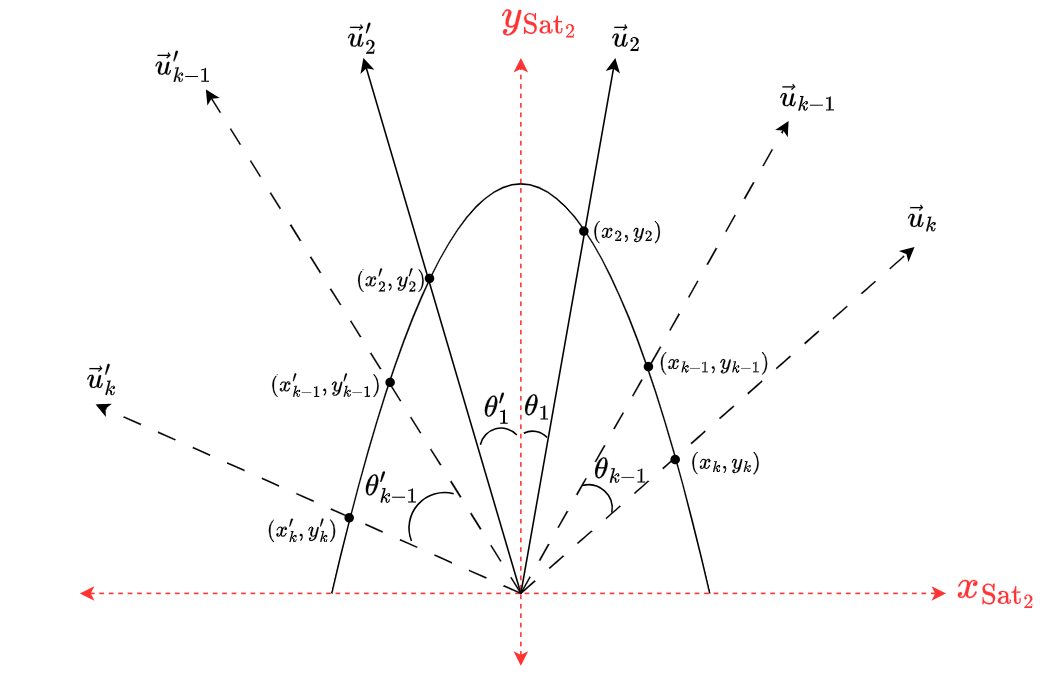}
    \caption{Finite spatial regions of asymmetric main beam radiation of $\text{Sat}_2$ on local frame where each $\vec{u}_i$ starts at the origin and $y_{\text{Sat}_2}$ is the beam direction axis.}
    \vspace{-5mm}
    \label{cap}
\end{figure}
\subsection{Thermal Noise Model}

According to black body radiation, any object in free space with non-zero temperatures radiate electromagnetic energy. Therefore, antennas are exposed to noise sources within their radiation fields. Other than man-made interference and disturbance sources on Earth, which can be foreseen in some extension, the free space of the solar system is full of unpredictable thermal emission bodies. 
In case the physical temperature and internal receiver losses of an antenna (e.g. transmission line, radiation efficiency) are neglected, the system temperature wıll be equal to the antenna temperature ($T_{sys} = T_A$) within a satellite.
\subsubsection {Cosmic Microwave Background (CMB) Radiation}
It is observed that CMB on the observable sky on ($\theta, \phi$) has non-uniform noise fluctuations with ensemble average temperature of $T^{CMB}=2.761 \pm 0.010$ K \cite{durrer2020cosmic} where the blackbody radiance peaks at $160.2$ GHz. The uncertainty of measurement sourced by instrument calibration, thermometer gain offset, and frequency offset \cite{fixsen2009temperature}. In free space, where the EM threats (e.g., atmospheric distortions, radiation belts, and geomagnetic storms) are minimized, it is appropriate to use high-gain transmitter and receiver amplifiers. For such a receiver front end, an unexpected power source is able to delude and even damage the receiver structure. As the CMB increasingly emerges as a dominant noise source on high-gain antennas, thorough analysis becomes imperative. 

\label{AA}
\subsubsection{Thermal Noise}
\label{therm}
A randomly located hidden celestial body to $p_c = [x_c, y_c]^T$ with a $d_C = ||p_c - p_{S_2}||_2$ distance to origin centered $\textrm{Sat}_2$ with $ p_{S_2}=[0,0]^T$ is a source of thermal noise with its brightness temperature $T_C$. Accordingly, the system temperature under effective thermal noise $(T_{eff})$ contribution from $N$ amount of celestial bodies is as follows: 
\begin{equation}
  T_{sys} = \sum_k {T_{eff}^k}, k \in \{1,2,..,N\}.
  \label{tsys}
\end{equation}
For celestial bodies with non-uniform brightness, $T_{eff}$ is the following:
\begin{equation}
T^{k}_{eff} = \frac{\int\limits_0^{2\pi} \int\limits_0^{\pi} T_C^{k}(\theta, \phi) G(\theta, \phi) \sin{(\theta)} d \theta d \phi}{\int\limits_0^{2\pi} \int\limits_0^{\pi} G(\theta, \phi) \sin{(\theta)} d \theta d \phi}, k\in\{1, 2,..,N\},
\end{equation}
where $G(\theta, \phi)$ is the receiver antenna gain. Noting that, $T^{CMB}_{eff} = T^{CMB}$ due to the independency of $\theta$ and $\phi$ in CMB. Similarly, in order to keep $T_C$ independent of $\theta$ and $\phi$, the thermal sky noise sourced by the celestial body regions are defined as following:
\begin{equation}
    \underset{x>0}{\textbf{T}_C}=\begin{cases}
		T_C^{1}, & \text{if $y > \frac{y_2}{x_2} \cdot x$}\\
            T_C^{2}, & \text{if $\frac{y_3}{x_2} \cdot x < y < \frac{y_2}{x_2} \cdot x$}\\
            \hspace{0.2cm}\vdots  & \hspace{0.5cm}\hdots \\
            T_C^{k}, & \text{if $y < \frac{y_{k+1}}{x_k} \cdot x$}
		 \end{cases}
   \label{eqth1}
\end{equation}
and 
\begin{equation}
    \underset{x<0}{\textbf{T}_C}=\begin{cases}
			T_C^{1}, & \text{if $y > \frac{y'_2}{x'_2} \cdot x$}\\
            T_C^{2}, & \text{if $\frac{y'_3}{x'_2} \cdot x < y < \frac{y'_2}{x'_2} \cdot x$}\\
            \hspace{0.2cm}\vdots  & \hspace{0.5cm}\hdots \\
            T_C^{k}, & \text{if $y < \frac{y'_{k+1}}{x'_k} \cdot x$}		 \end{cases}
   \label{eqth2}
\end{equation}
where $T_C^{i} \in {\textbf{T}_C}$ is the thermal noise on the corresponding radiation region of receiver occupancy, respectively. As shown in Figure \ref{cap}, each existing region corresponds to a noise temperature source. For the continuity on $\textbf{T}_C$, Eq. (\ref{eqth1}) and Eq. (\ref{eqth2}) assume that antenna nulls does not exist. Due to the directive radiation, thermal background noise occupancy regions caused by the celestial bodies denoted with beam efficiencies (BE) for each spatial region are given by: 
\begin{equation}
\text{BE}^i = \frac{\int\limits_0^{2\pi} \int\limits_0^{\theta_i} U(\theta, \phi) d \theta d \phi}{\int\limits_0^{2\pi} \int\limits_0^{\pi} U(\theta, \phi) d \theta d \phi},\hspace{0.5cm} i\in\{1, 2,.., k-1\},
\end{equation}
where each spatial region is separated with $\vec{u}_i \in \mathbb{R}^2$. $\theta_i \in (-\pi,\pi]$ is the beamwidth of the corresponding region such as $ \vec{u}_i \cdot \vec{u}_{i-1} = |\vec{u}_i| |\vec{u}_{i-1}| \cos(\theta_{i-1}), \forall i$ and $U(\theta,\phi)$ is the radiation intensity. Thus, $T_{sys}$ can be formulated as linear superposition of $\text{BE}^{i}, \forall{i}$. Herewith, $T_{sys}$ in Eq. (\ref{tsys}) can be written explicitly as the following:
\begin{align}
    T_{sys} &= \sum_i \text{BE}^i \times T^i_{C}\\ 
    &= \text{BE}^{1} \times T^1_{C} + \text{BE}^{2} \times T^2_{C}+ \cdots \text{BE}^{k} \times T^k_{C} + T^{CMB}.
\end{align}
Consequently, $T_{sys}$ is obtained under both known and unknown stellar bodies for Earth-to-Moon IP-ISL. 
\subsection{Channel Model}
% Under the deep space channel assumption, the atmospheric effects on ISL are neglected. Therefore, weak considered noise sources becomes predominant on free space such as CMB, hardware noise, celestial body radiations and solar activities.  This positioning serves to effectively reduce localization and PAT errors, which were assumed to be negligible in this study. 
Wave propagation is primarily influenced by solar activity, resulting in phase fluctuations during transmission. Solar winds release charged particles into space at velocities ranging from $400$ to $800$ km/s, existing in a plasma state with energies between $0.5$ and $10$ keV. The Sun-Earth probe (SEP) angle $\alpha$ determines the distribution of charged particles in relation to the real permittivity $\epsilon$. Starting the channel definition with Drude-Lorentz model, which can describe the harmonic oscillation of the released particles with dielectric function as following
\begin{equation}
\label{loren}
    \epsilon = 1 + \frac{\omega_p^2}{\omega_0^2 -\omega^2},
\end{equation}
where $\omega_p$ is the angular plasma frequency, $\omega$ is the angular
frequency of the Lorentz oscillator and $\omega_0$ is the resonance frequency. The plasma frequency ($f_p$) is defined as following
\begin{equation}
    f_p = \frac{e}{2 \pi} \sqrt{\frac{n_e}{\epsilon_0 m_e}},
\end{equation}
where $e$ is the electron charge, $n_e$ is electron density, $m_e$ is the effective mass of the electron, and $\epsilon_0$ is the vacuum permittivity. Assuming that the interaction of charged particles with each other does not exist, the open expression of $\epsilon$ is the following:
\begin{equation}
    \epsilon = 1 + \frac{\lambda^2 n_e e^2}{\epsilon_0 m_e (\lambda^2 \omega_0^2 - 4 \pi^2 c^2)}.
    \label{eps2}
\end{equation}
The deep space weather of solar wind with dynamic electron density model is \cite{muhleman1977electron}:
\begin{equation}
    n_e(r) = \frac{2.21 \times 10^{14}}{(r/r_s)^6}+\frac{1.55 \times 10^{12}}{(r/r_s)^{2.3}} \hspace{0.5cm} [m^{-3}],
    \label{densdyn}
\end{equation}
where $r_s$ is the Sun radius and $r$ is the perpendicular distance from Sun vertex to the shortest $\textrm{Sat}_1$ - $\textrm{Sat}_2$ path. Here, the plasma density is dominated by the first term of Eq. (\ref{densdyn}) and has a radial spread around the Sun \cite{ho2008solar}.

%Accordingly, the positions of each satellites are $p_{S_1}$ and $p_{S_2}$. Thence, $d_S = ||p_{S_1}-p_{S_2}||$.

High plasma frequency draws $\epsilon$ away from being $1$, leading to both phase and angular fluctuations ($\delta(\theta)$) along the wave trajectory. The phase shift ($S$) on IP-ISL is $S = S_0 + \varphi$ where $S_0$ is the nominal phase shift when $\omega_p = \omega$, and $\varphi$ is the random variable of phase shift\footnote{Result of $\varphi$ on propagating wave also leads to offsets on the carrier frequency (e.g. Doppler noise) which is neglected on this study.} caused by $\Delta \epsilon$, which corresponds to variant terms in Eq. (\ref{loren}). The nominal phase shift is supposed to be estimated and compensated by $\text{Sat}_2$, and considered as $S_0 = 0$. Letting $\epsilon(\ell)$ be unitary constant along the path, and the integration path ($d_S$) to be invariant and independent of $\epsilon$. Accordingly, $\varphi$ is the following:
\begin{align}
    \varphi &= \frac{1}{2}k \oint \frac{1}{\sqrt{\epsilon(\ell)}} \Delta \epsilon d\ell,\\
    &\approx \frac{k}{2} \int^{d_S}_0 \Delta \epsilon d\ell.
    \label{varphii}
\end{align}
wherein $\Delta \epsilon$ denotes the deflections due to non-constant reflectivity along the path. Recall that $\epsilon$ is a differentiable function, thus; $\Delta \epsilon$ can be expressed from Eq. (\ref{eps2}) as following
\begin{equation}
    \Delta \epsilon = \frac{(\lambda e)^2}{\epsilon_0 m_e (\lambda^2 \omega_0^2 - 4\pi^2 c^2)} \Delta n_e.
    \label{deltae}
\end{equation}
The irregularities of plasma density on high-altitude orbits are modeled proportionally with the magnitude of plasma density,
\begin{equation}
    \Delta n_e \approx \frac{n_e}{\alpha \beta}.
\end{equation}
where $\beta$ denotes the order of solar plasma magnitude, and $\alpha$ is the electron density index as a scaling factor. $\Delta n_e$ is not monotonically decreasing, and the model is viable as long as $(r/r_s)^6 >> \alpha \beta$. Lastly, the Chandrasekhar equation notates the $\delta \theta$ with orthogonal phase shift component to the wavefront as following
\begin{equation}
\label{chad}
    \delta \theta = \frac{1}{k\sqrt{\epsilon}} \nabla_{\bot} \varphi.
\end{equation}
With this statement, the interplanetary IP-ISL channel model is completed.
\subsection{Angular Fluctuations Analysis}
We present a closed form expression of angular fluctuation variance $\mathbb{E}[\delta \theta ^2]$ analysis on $\textrm{Sat}_1-\textrm{Sat}_2$ link disturbed by solar plasma density as illustrated in Figure \ref{ISLGeo}. Let $\epsilon=1$, angular fluctuation in reflectarray with aperture area $A$ along the $x$ axis unit cells is following \cite{wheelon2001electromagnetic}
\begin{equation}
    \delta \theta = \frac{1}{kA} \int \int_A \frac{\partial}{\partial x} \varphi dx dz.
    \label{eq17}
\end{equation}
For the remaining analysis, isotropic assumption will be made along the $z$ axis. Using  Eq. (\ref{varphii}) and Eq. (\ref{deltae}), $\varphi(x,z)$ is encountered for each eddy $\ell$ is the following
\begin{align}
    \varphi(x,z) &= \frac{k}{2} \int^\infty -\lambda^2 r_e \Delta n_e d\ell,\\
    &= \pi \lambda r_e \int^\infty - \Delta n_e d\ell,
    \label{varphixz}
\end{align}
where the radius of electron is used as $r_e = \frac{e^2}{4 \pi c^2 \epsilon_0 m_e }$. In the subsequent equations, $\delta_ \theta$ is expressed using Eq. (\ref{varphixz}) applied to Eq. (\ref{eq17}), and 
\begin{equation}
   \delta \theta =  -\frac{1}{2A} r_e \lambda^2 \int \int_A \frac{\partial}{\partial x} \Bigl(\int^\infty \Delta n_e d \ell\Bigl) dx dz,
\end{equation}
\textcolor{black}{is obtained}. Instead of applying Leibniz integral rule directly, the derivative of the characteristic function $\xi_X (t)$ is used with the assumption that $n$-th moment exists such that $\mathbb{E}[X^n]~=~i^{-n}\Bigl[\frac{\partial ^n}{\partial t^n} \xi_X (t) \Bigl]_{t=0}$ as following,
%\begin{multline}    
%    \mathbb{E}[\delta \theta ^2] = \frac{1}{4A^2}r_e^2 \lambda^4 \int \int_A dx dz \int \int_A \frac{\partial^2}{\partial x \partial x'} \\
%    \times \mathbb{E}[\Delta n_e (x,z) \Delta n_e (x',z')] dx' dz'.
%    \label{eq21}
%\end{multline}
\begin{align}
    \mathbb{E}[\delta \theta ^2] &= \frac{1}{4A^2}r_e^2 \lambda^4 \int \int_A dx dz \int \int_A dx' dz' \int^\infty_0 d \ell \nonumber \\ &\times \int^\infty_0  \frac{\partial^2}{\partial x \partial x'} \mathbb{E}[\Delta n_e \Delta n_e'] d \ell '   \label{ewq21}
%    &= \frac{1}{4A^2}r_e^2 \lambda^4 \int \int_A dx dz \int \int_A dx' dz' \\
%    &\times \int \int_A \frac{\partial^2}{\partial x \partial x'} \mathbb{E}[\Delta n_e (x,z) \Delta n_e (x',z')] dx' dz'. \nonumber
\end{align}
where the symmetry order of the derivatives is held by Clairaut's theorem \cite{aksoy2002mixed}. The spatial correlation of electron density variation on Eq. (\ref{ewq21}) is shown with the wave number spectrum \cite{ivarsen2023measuring} that utilizes the electron density fluctuation spectrum ($\Psi_N$) as follows:
\begin{equation}
    \mathbb{E}[\Delta n_e  \Delta n_e'] = \int \Psi_N(\kappa) e^{i \kappa_x (x-x') + i \kappa_z (z-z')  + i \kappa_\ell (\ell - \ell ')} d^3\kappa,
\end{equation}
where $\kappa$ notates the turbulence wave number and resolves with the following:
\begin{align}
\label{var1}
        \mathbb{E}[\delta \theta ^2] &= \underbrace{\frac{1}{4A^2}r_e^2 \lambda^4 \int \int_A dx dz \int \int_A  e^{i \kappa_x (x-x') + i \kappa_z (z-z')} dx' dz'}_{\text{\textbf{I}}}   \\ &\times \underbrace{\int \kappa^2 d^3 \kappa   \int^\infty_0 d \ell \int^\infty_0  \Psi_N(\kappa)  e^{i \kappa_\ell (\ell - \ell ')} d \ell '}_{\text{\textbf{II}}} \nonumber
\end{align}
Accordingly, the elliptic reflectarray surface integration on Eq. (\ref{var1})-\textbf{I} in polar coordinates is given as: 
\begin{align}
    \text{\textbf{I}} = &\frac{r_e^2 \lambda^4}{4\pi^2a^2b^2} \int^{r} abr_{mn}^{(1)} dr_{mn}^{(1)} \int^{r} abr_{mn}^{(2)} dr_{mn}^{(2)} \\ &\times  \int^{2\pi} d\Theta_1  \int^{2\pi} d\Theta_2  \\
    &\times  \exp{\bigl[i \kappa_x \bigl(r_{mn}^{(1)}\cos{(\Theta_1)}-r_{mn}^{(2)}\cos{(\Theta_2)}\bigl)\bigl]} \nonumber \\
    &\times \exp{\bigl[i \kappa_z \bigl(r_{mn}^{(1)}\sin{(\Theta_1)}-r_{mn}^{(2)}\sin{(\Theta_2)}\bigl)\bigl]} \nonumber,
\end{align}
where $x = r\cos{(\Theta)}$, $z = r\sin{(\Theta)}$ and  $r_{mn}$ is the distance of the $mn^{\text{th}}$ unit cell from the center of the aperture. Using the zeroth-order Bessel function of the first kind ($J_0$) on definite integral \cite{watson1922treatise}-[Ch.2, pg. 20] $J_0(z) = \frac{1}{2\pi} \int^{2\pi} e^{(z\cos{(x)})}dx$ and $u \cos{(x)}+v\sin{(x)} = \sqrt{u^2+v^2} \cos{(x+\Theta)}$ for some $\Theta$: 
\begin{align}
    \text{\textbf{I}} = &\frac{r_e^2 \lambda^4}{a^2b^2} \int^r abr_{mn}^{(1)}  J_0(r_{mn}^{(1)} \sqrt{\kappa_x^2 + \kappa_z^2}) dr_{mn}^{(1)} \\ &\times \int^r abr_{mn}^{(2)} J_0(r_{mn}^{(2)} \sqrt{\kappa_x^2 + \kappa_z^2})  dr_{mn}^{(2)} \nonumber. \\
     &= \frac{r_e^2 \lambda^4}{a^2 b^2 r^2} \biggl[\frac{J_1(r \sqrt{\kappa_x^2 + \kappa_z^2})}{r \sqrt{\kappa_x^2 + \kappa_z^2}} \biggl]^2.\nonumber
\end{align}
Due to the fact that $d_S >> \ell$, arclength integration can be done over $y$-axis in Eq. (\ref{var1})-\textbf{II} as following:
\begin{equation}
    \text{\textbf{II}} = 2 \pi \int \kappa^2 \Psi_N(\kappa)  \delta(\kappa_\ell) d^3 \kappa    \int^\infty_0 d y. 
\end{equation}
By this, the closed form expression for angular error variance can be obtained from Eq. (\ref{var1}). In further steps, expanding the final expression with $\Psi_N$ reveals the contribution of electron density on the angular fluctuation.
%\begin{figure}[] 
 %   \centering
%    \includegraphics[width=0.5\textwidth]{sat1sat22.png}
%    \caption{The illustration for diffracted trajectory along the $\textrm{Sat}_1-\textrm{Sat}_2$ path due to $\Delta \epsilon$, leading to $\delta\theta$ deviation on direction  of arrival (DoA).} 
    %\vspace{-5mm}
 %   \label{cap}
%\end{figure}
\subsection{Signal Model}
\begin{figure}[] 
      \includegraphics[width=0.50\textwidth]{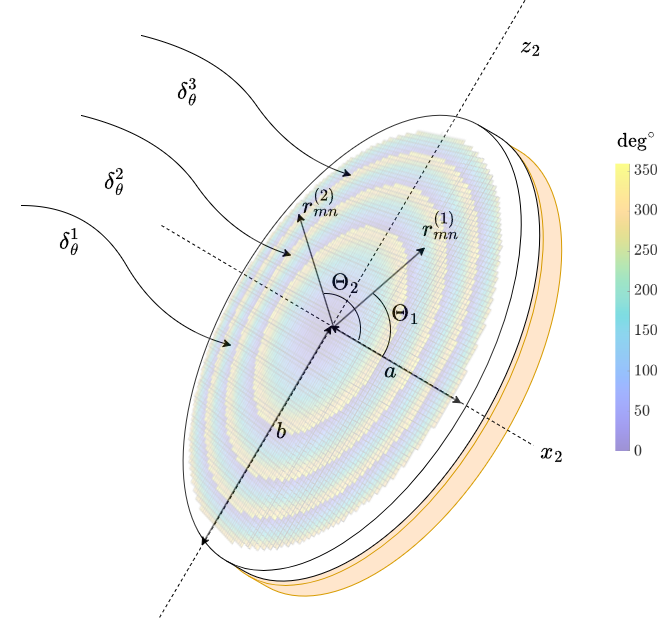}
      \vspace{5mm} 
    \caption{Angle of arrival fluctuations arise from the non-constant dielectric of the \textcolor{black}{free space} plasma within a time instance of the reflectarray aperture phase distribution.}
    %\vspace{-17mm}
    \label{geo}
\end{figure}
The transmission of average signal power of $\sqrt{P}$ by $\textrm{Sat}_1$ having $M_1\times M_2$ unit cells in the reflectarray on $xz-$ plane to $\textrm{Sat}_2$ having an identical reflectarray, received signal model is the following:
\begin{equation}
    y =  \sqrt{P} h( s + \eta_t) + \eta_r + w
\end{equation}
where $h=\sqrt{\mu}e^{j\varphi}$ and $h \in \mathbb{C}$ is the effective channel coefficient with path gain $\mu \sim (\frac{\lambda}{4 \pi d_s})^2$ for the sample of $s \in \mathbb{C}$, $\eta_t$ and $\eta_r$ are aggregated hardware distortions on the transmitter and receiver respectively with $\eta_t \sim \mathbb{CN}(0, P\kappa_1^2)$ and $\eta_r \sim \mathbb{CN}(0, \kappa_2^2 P |h|^2)$ where $\kappa_1, \kappa_2 \geq 0$ represents the independent hardware noise levels with aggregated distortion at the receiver with 
\begin{equation}
\mathbb{E}[|h\eta_t + \eta_r|^2]=P|h|^2(\kappa_1^2 + \kappa_2^2),    
\end{equation}
and $w$ is the accumulated system thermal noise analyzed on Section \ref{therm} with power spectral density of $N_{sys}/2$. In this case, the independent, effective distortion noises are $\eta_t$ and $\eta_r/h$, which follow Gaussian distribution \cite{bjornson2013new}. This is one reason why the Earth orbiter $\text{Sat}_1$ is selected as a transceiver, as the RF impairments are reliably monitored and easily replaceable. Thence, the spectral efficiency ($\textrm{SE}$ [bp/sec/Hz]) of the corresponding link with polarization and alignment matched reflectarrays is the following:
\begin{equation}
    \textrm{SE} = \log_2{(1 + \frac{P G_T(\theta,\phi) G_R(\theta+\delta \theta, \phi) |h|^2}{N_{sys} + (\kappa_1^2 + \kappa_2^2)P})},
\end{equation}
where $N_{sys} = kT_{sys}$, and $k$ is the Boltzmann constant. 
%with Chandrasekhar equation is following:
%\begin{equation}
 %   \delta(\theta)=\frac{1}{k\sqrt{\epsilon}} \nabla_\perp \Phi
%\end{equation}
\section{ISL with Reflectarray Antenna}
\label{sec3}
A reflectarray consists of unit cells over an aperture to reflect the incoming signal. A planar reflectarray with geometric setup on Figure $\ref{geo}$ is used for the Earth-to-Moon IP-ISL application. The system consists of an offset reflectarray with the feed (in our case, a horn antenna) located at the focal distance. The reflectarray is an ellipse with $M_1 \times M_2$ unit cells, and its projected aperture in the $zy$- plane is a circle with radius DMZ-XMC/2. The phase distribution of the reflectarray manipulates the incident beam to reflect it towards the direction $\theta_s$.  

\subsection{Reflectarray}
\begin{figure}[b] 
    \label{geo}
     \includegraphics[width=0.5\textwidth]{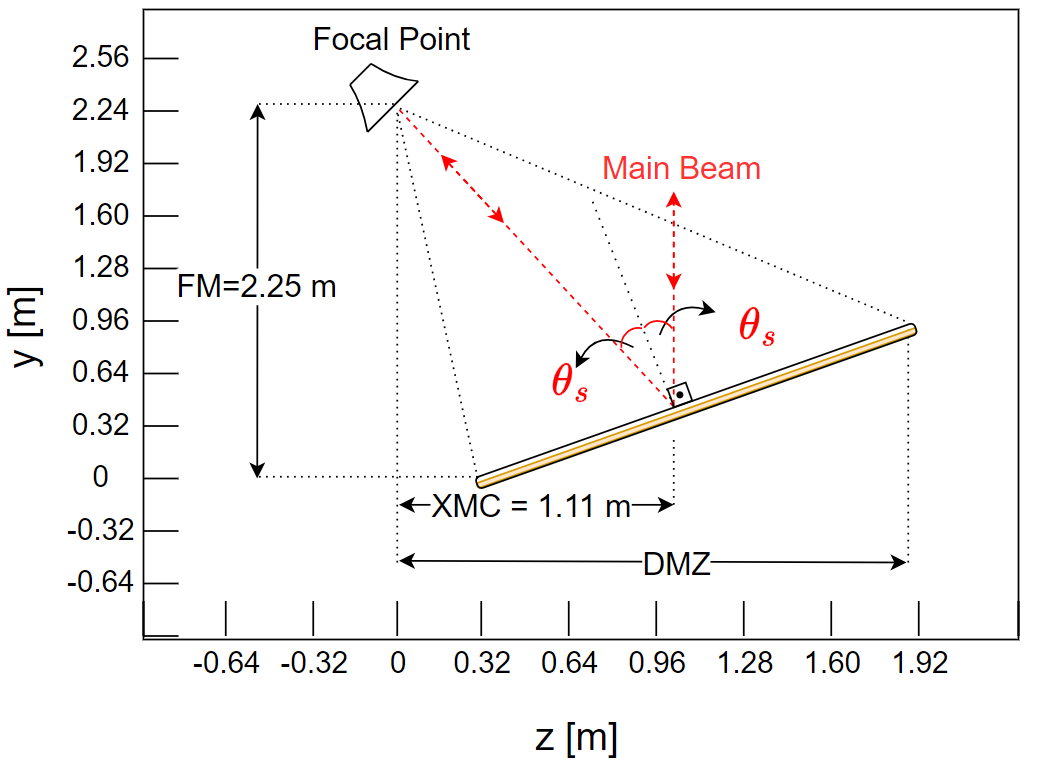}
    \caption{Planar reflectarray geometry design for both $\text{Sat}_1$ and $\text{Sat}_2$.}
    %\vspace{-5mm}
    \label{geo}
\end{figure}
The radiation pattern for the reflectarray, wherein the phase delay of the $mn^{\text{th}}$ unit cell ($\varphi_{mn}$) is assigned in order to direct the main beam towards the direction $\hat{u}_0$, is obtained using the following \cite{nayeri2018reflectarray} - pg.82: 
\begin{multline}
      E(\theta, \phi) = \sum_{m=1}^{M} \sum_{n=1}^{N} \frac{\cos^{q_f}{(\theta_f (m,n))}}{|r_{mn}-r_f|} \\ \times  e^{-jk(|r_{mn}-r_{f}|-r_{mn} \cdot \hat{u})} \cos^{q_e}{\theta_e(m,n)} e^{j \varphi_{mn}}.  
\end{multline}
where $\theta_f$ denotes the elevation angle in feed's coordinate system, $r_f$ and $r_{mn}$ are the position vectors of the feed and the $mn^{\text{th}}$ unit cell, and $\cos^{q_f}(\cdot)$ is the balanced feed radiation pattern.

\subsection{Beamsteering Resolution}
%Steering angle ($\theta_s$) will result in scan losses, causing a wider main beam and sidelobes as $\theta_s > \theta_\circ$ where $\theta_\circ$ is the arbitrary tolerance angle for the steering range of the antenna. 
 
\textcolor{black}{We denote the required beamsteering angle pair with $(\theta_s, \phi_{s})$ such that ($\theta_s, \phi_s) := (\delta \theta_x,\delta \theta_z$), which fully compensates AoA fluctuation. Due to a scan loss tolerance in beamsteering, the compensation is limited with a predefined design parameter $\theta_0$ such that $\theta_{s},\phi_s \in (-\theta_0,\theta_0]$. Therefore, $\theta_{s}$ and $\phi_s$ are intentionally limited with $\theta_0$ at most.
%Thus, $\theta_s > |\theta_0|$ leads to a wider main beam and sidelobes
\begin{figure}[] 
    \centering
    \includegraphics[width=0.5\textwidth]{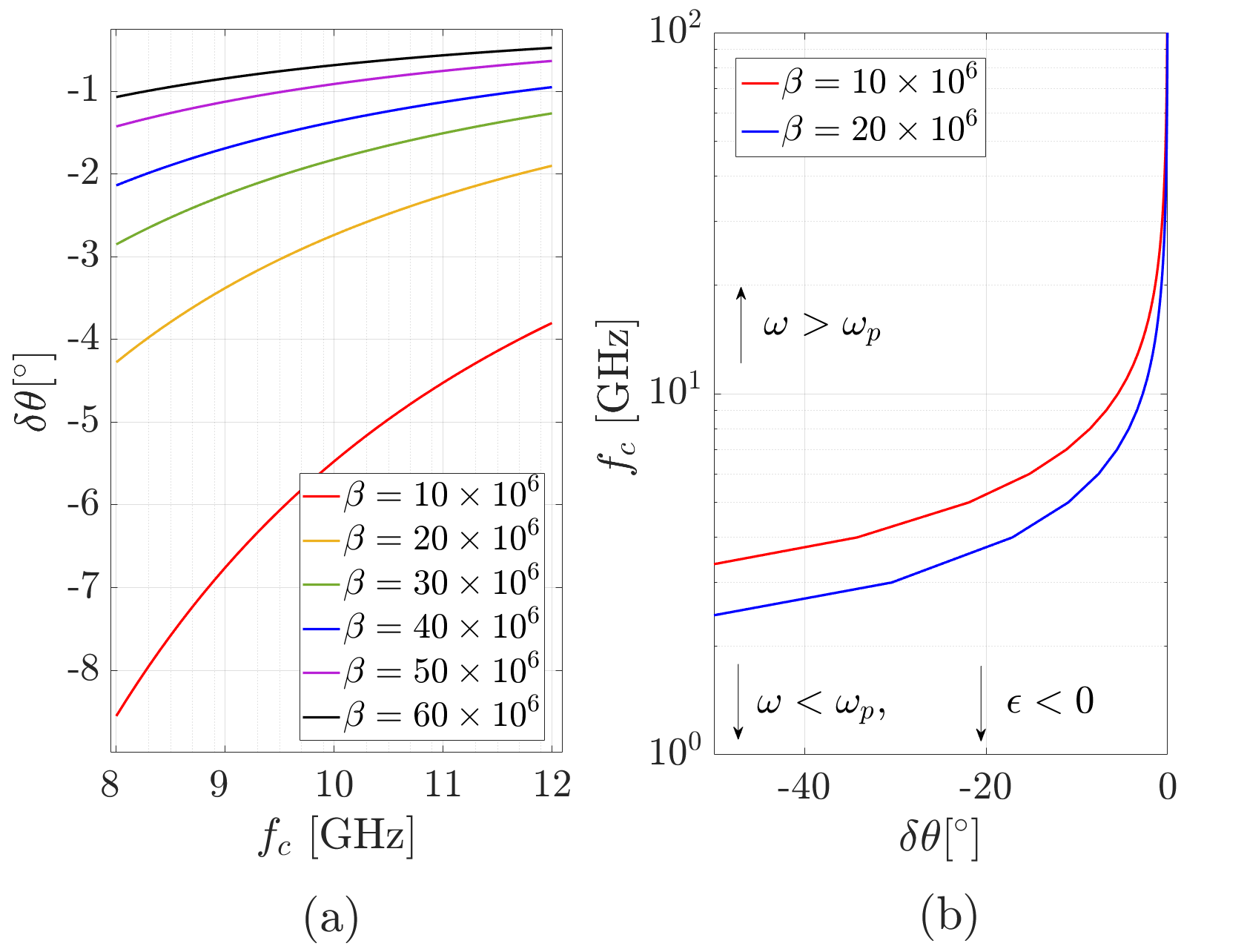}
    \caption{(a) The vulnerability to diffractions on the carrier within different solar plasma levels. (b) Robustness of the high-frequency band against X-band in solar plasma density.}
    %\vspace{-5mm}
    \label{xvsthz}
\end{figure}
In an ideal case with perfect estimation of $\varphi$ for each instance of time, $(\delta \theta_x, \delta \theta_z)$ loss can be compensated by $(\theta_s, \phi_s)$ if and only if $-\theta_{0} \leq \delta \theta \leq \theta_{0}$. The required phase profile on the aperture of the $\text{Sat}_2$ reflectarray in order to direct the beam to $(\theta_s, \phi_s)$ is calculated using the following formula \cite{nayeri2018reflectarray}-Section 2.4:
\begin{equation}
    \phi_A = k_0 \left(R_i - \sin{\theta_s} \left(x_i \cos \phi_s + z_i \sin \phi_s\right) \right) + \varphi_0,
\end{equation}
where $k_0$ is the wave number, and $R_i$ is the distance between the $i^{\text{th}}$ unit cell and the phase center of the feed antenna. $x_i$ and $y_i$ refer to the coordinates of the $i^{\text{th}}$ unit cell on the aperture of the reflectarray.} The first term in this formula serves to compensate for the nonuniform phase of excitation received from the feed antenna, while the terms dependent on $(\theta_s,\phi_s)$ apply the required phase profile for beamsteering. One another advantage of the reflectarray is that it can be used for the estimation of $\delta \theta$ without requiring knowledge of the state of the $\text{Sat}_1$. 
%Note that, strong external heat radiations will challange and leads to following optimization problem: 
%\begin{multline}
%    \textrm{minimize} {(T_{sys},\varphi)\\
%   \text{s.t.} -\theta\circ \leq \theta_B \leq \theta_\circ
%\end{multline}
\begin{figure}[] 
    \centering
    \includegraphics[width=1\linewidth]{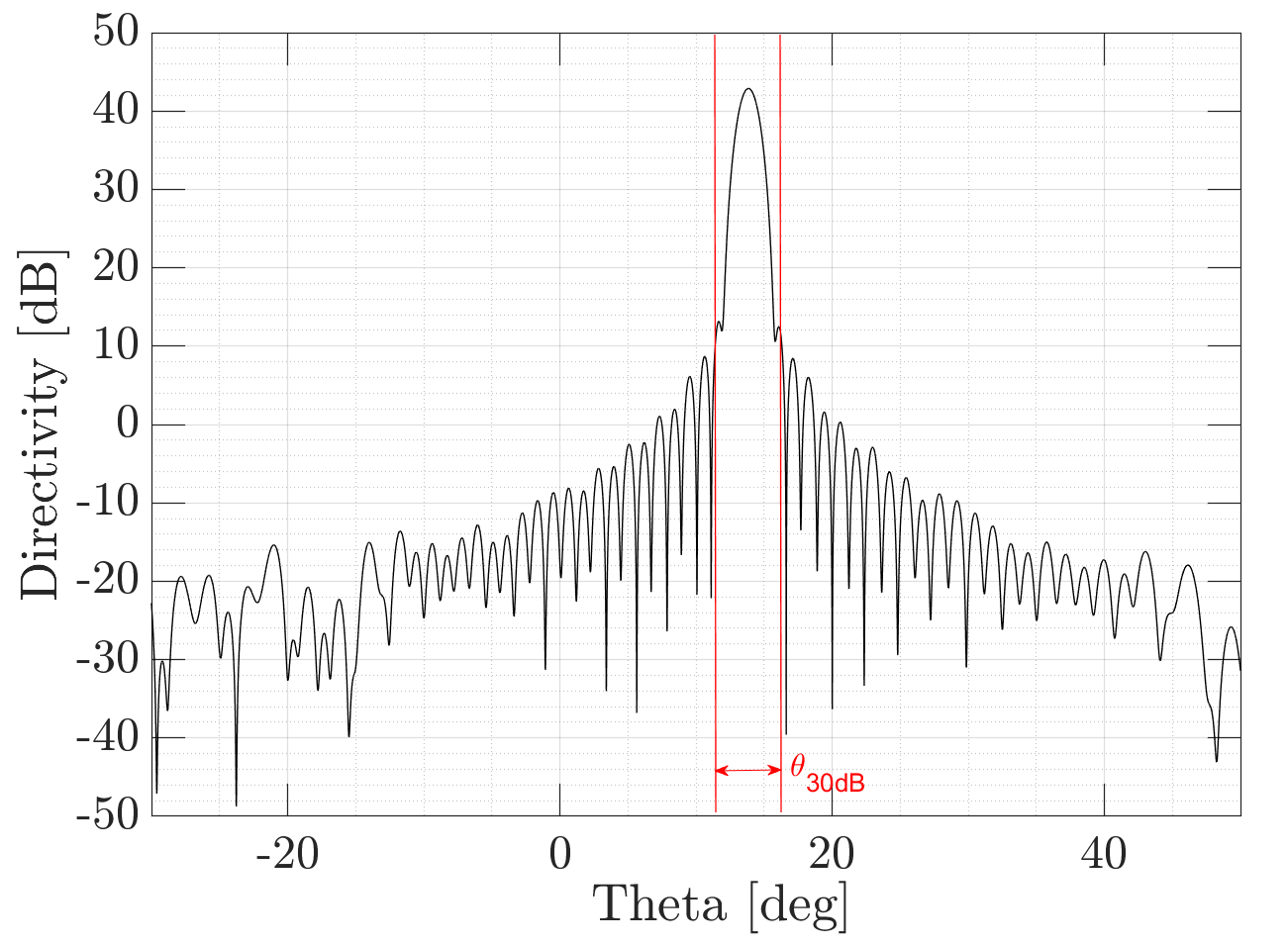}
    \caption{Directivity of the designed planar reflectarray with $\theta_{ref}$ wave angle.}
    %\vspace{-5mm}
    \label{direc}
\end{figure}
\begin{figure*}[] 
    \centering \includegraphics[width=0.8\linewidth]{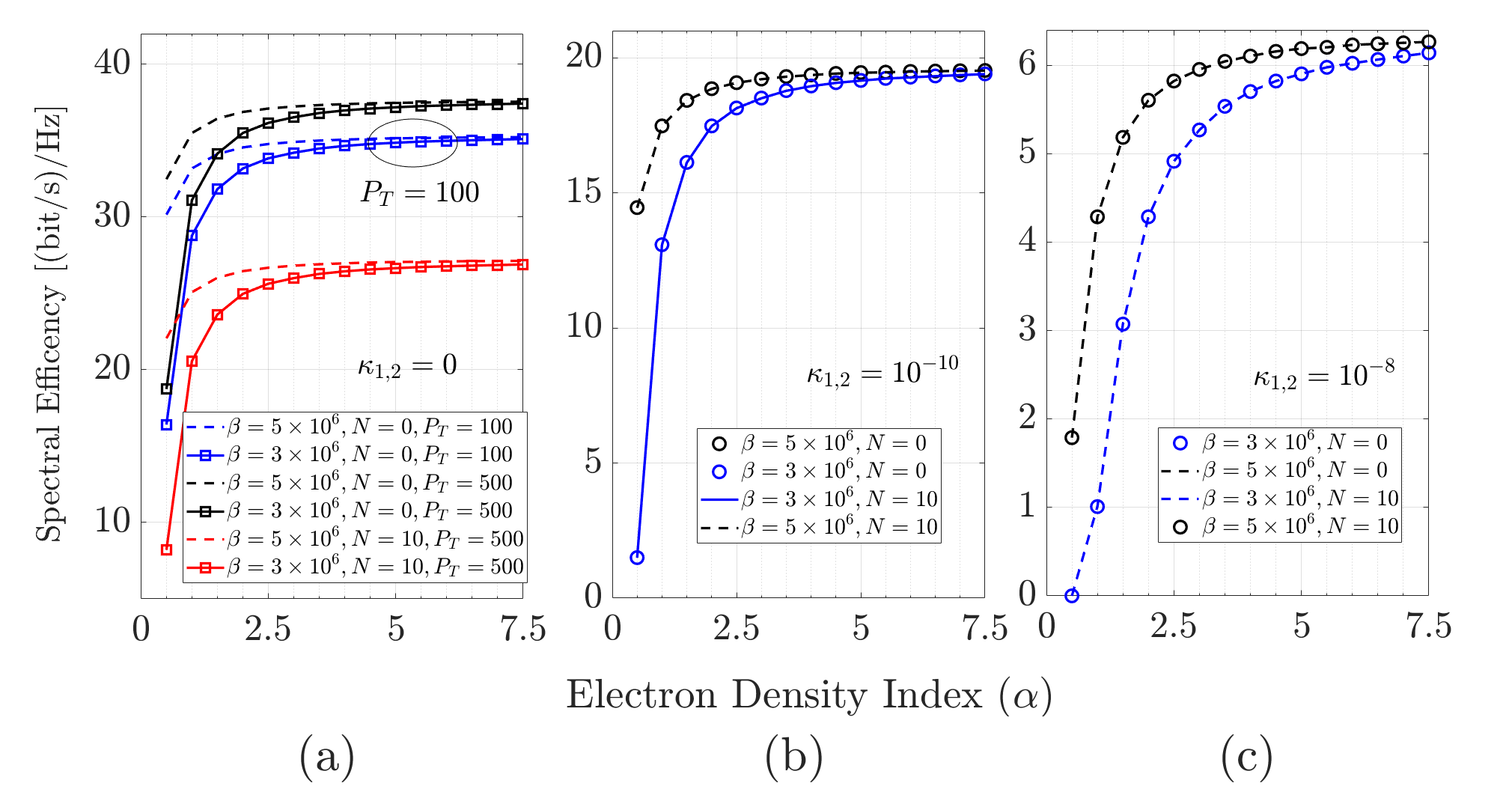}
    \caption{Reflectarray use on different $\kappa$ cases within $\beta$ solar plasma indexes under $N$ celestial bodies.}
    %\vspace{-5mm}
    \label{betase}
\end{figure*}
\vspace{-5mm}
\begin{figure*}[] 
    \centering \includegraphics[width=0.8\linewidth]{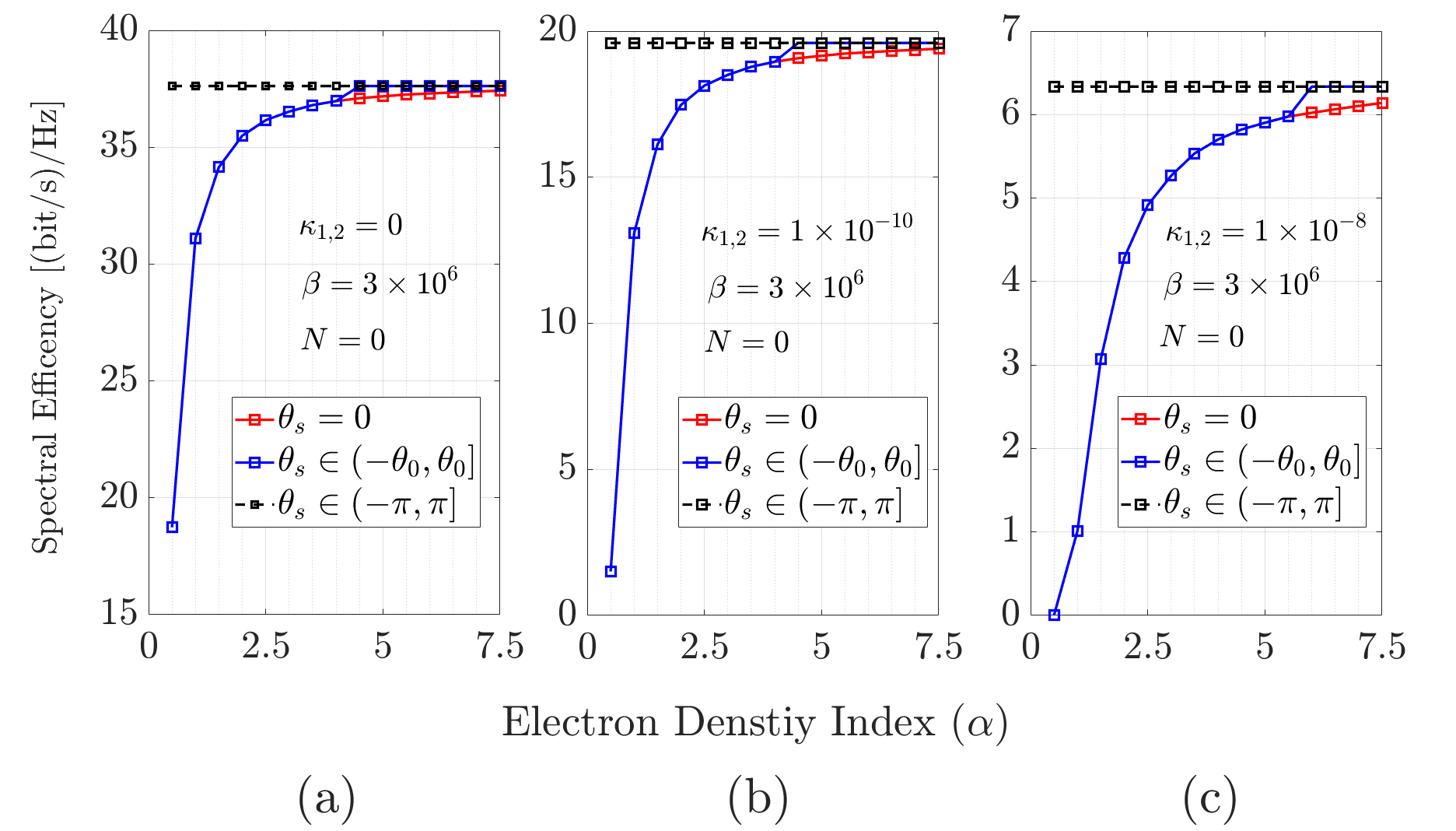}
    \caption{Expected spectral efficiency with ideal beamsteering of $\theta_s$ without scan loss.}
    %\vspace{-5mm}
    \label{refse}
\end{figure*}
\section{Numerical Results}
\label{sec4}
IP-ISL performance evaluation is conducted over numerical simulations. $\text{Sat}_1$ to $\text{Sat}_2$ transmission is investigated on deep space channel under $N$ randomly located number of celestial bodies\footnote{We assume celestial bodies not to obstruct the line of sight IP-ISL; instead, they are solely regarded as sources of thermal noise.} on the main beam field-of-view $\theta_{\textrm{30dB}}$. The satellites possess $\kappa = \kappa_{1,2}$ hardware impairments. Despite the zero misalignment, the channel is exposed to $\delta_\theta$ fluctuations on the $\text{Sat}_2$. Assuming that the IP-ISL coverage is on the half sphere of each planet due to the tidal locking and the circular shape of orbits, the closest distance is $d_s$ and longest distance is $\sqrt{(d_s + h_1 + h_2)^2 + (h_1+h_2)^2}$ wherein $\text{Sat}_1$ and $\text{Sat}_2$ are at the opposite edge end of the coverage. Therefore, the $T$ sample average of SE between varying distances is considered the expectation of the SE for IP-ISL. 

Several different methodologies are used to realize this analysis. The following section explains this strategy.
\subsection{Numerical Methodologies}
The directivity of an antenna ($D$) in the ($\theta_0,\phi_0$) direction can be expressed as  
\begin{equation}
    D = \frac{4 \pi |E(\theta_0,\phi_0)|}{\int\limits_0^{2 \pi} \int\limits_0^\pi |E(\theta,\phi)|^2 \sin{\theta} d\theta d\phi}.
\end{equation}
The directivity of the reflectarray is evaluated numerically with series approximation over $N$ uniform divisions of $\pi$ intervals \cite{balanis2016antenna} and illustrated in Figure \ref{direc}. Similarly, the numerical integration for the Chandrasekhar equation is solved by iterative rectangular rule, with the discretized number of steps of $1000$ \cite{griffiths2006numerical}. The gain of the antennas depend on the constant antenna efficiencies $\nu_t$ and $\nu_r$. The design parameters of the identical reflectarrays on both transmitter and receiver side are given in Table \ref{reftable}. 
\subsection{Discussion}
Figure \ref{betase} shows the impact of the different solar plasma order values $\beta$, hardware impairments $\kappa_{1,2}$, transmission power $P_T$, and the number of random celestial bodies $N$ on the IP-ISL system. The first sub-figure shows that the ideal $\textrm{Sat}_1$ and $\textrm{Sat}_2$ design with $\kappa_{1,2}=0$ reveals the substantial impact of celestial bodies on SE, especially on low electron density index $\alpha$ regions. For the all simulations, $P_T = 500$, except for that shown in Figure \ref{betase}-a. On Figure \ref{betase}-a, randomly sized $N=10$ celestial bodies spread on $\theta_{\textrm{30dB}}$ beamwidth of $\textrm{Sat}_2$ limits the maximum SE with more than $10~\textrm{(bit/s)/Hz}$. On the other hand, $P_T=100$ and $P_T=500$ differ only when $\alpha<2.5$. This shows that the unmitigated $\delta_\theta$ bottlenecks the IP-ISL and increment in $P_T$ simply is not enough to achieve higher SE. Another interesting result on Figure \ref{betase}-b and Figure \ref{betase}-c is that the $N=0$ case coincides with $N=10$ for each $\beta$. This shows that celestial body impact over the system is suppressed by the hardware impairments. In a realistic IP-ISL system, this implies that a certain number of hidden celestial bodies can be disregarded as noise sources due to the existing hardware impairments. Three subfigures show that an SE floor exists for each IP-ISL design, even if there are no hardware impairments. 
\begin{figure}[t] 
    \centering
\includegraphics[width=0.49\textwidth]{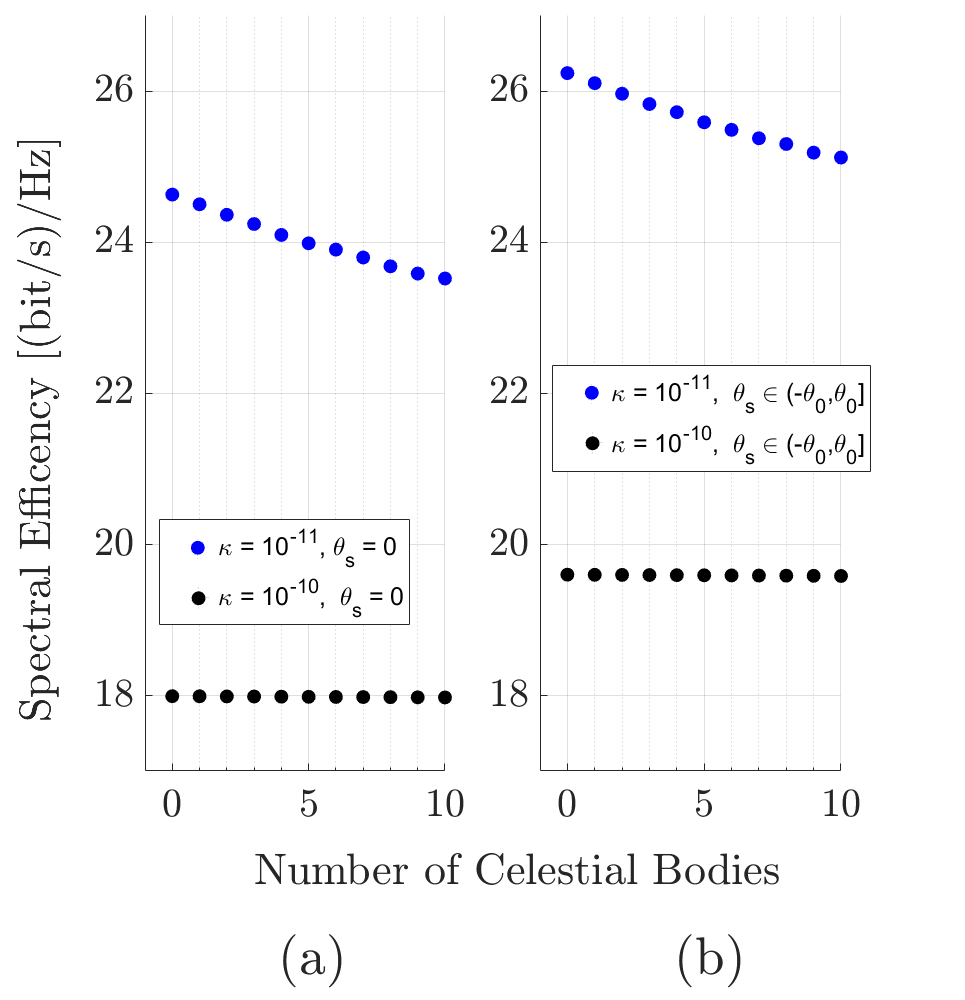}
    \caption{Suppressed celestial body source noise with high $\kappa_{1,2}$ impairments.}
    %\vspace{-5mm}
    \label{celes1}
\end{figure}
In Figure \ref{refse}, the use of ideal reflectarray with different $\theta_s$ is given for constant solar plasma order $\beta$. Expectedly, an ideal beamsteering scenario within $(-\pi,\pi]$ brings the IP-ISL to its maximum spectral efficiency. With a limited $\theta_0$ steering angle, there exists an $\alpha$ index where the reflectarray can fully recover the SE loss. Three cases with different $\kappa$ indicate that the full recovery point gets further away as the hardware impairments increase. Thus, the existing $\kappa$ bottlenecks the reflectarray beamsteering utilization. Nevertheless, beamsteering gain exists even in the high hardware impairment cases. Comparing the Figure \ref{refse} with Figure \ref{betase}, the ideal reflectarray beamsteering gain can go up to $30~\textrm{(bit/s)/Hz}$. However,in a limited yet realistic $\pm \theta_s$ steering interval, the SE can be increased by $\sim 1~\textrm{(bit/s)/Hz}$. This reveals the importance of the \textcolor{black}{steering range} $\theta_s$ of the reflectarray. Interestingly, $\sim 1~\textrm{(bit/s)/Hz}$ SE gain is not limited by the hardware impairments, only the solar plasma density. 

Figure \ref{celes1} illustrates the impact of the number of celestial bodies $N$ on IP-ISL for a sample averaged $\alpha$ from $\alpha = [1, 10]$ and constant $\beta = 3 \times 10^{-10}$. Both figures verify the previous inference that the celestial bodies are negligible for $N<10$ for $\kappa>10^{-10}$ cases. Beamsteering is not applied in the first sub-figure Figure \ref{celes1}-a. On the Figure \ref{celes1}-b, $\theta_0$, beamsteering achieves $\sim 1~\textrm{(bit/s)/Hz}$ for every $\kappa$ and $N$. 

\begin{figure}[t] 
    \centering
    \includegraphics[width=0.51\textwidth]{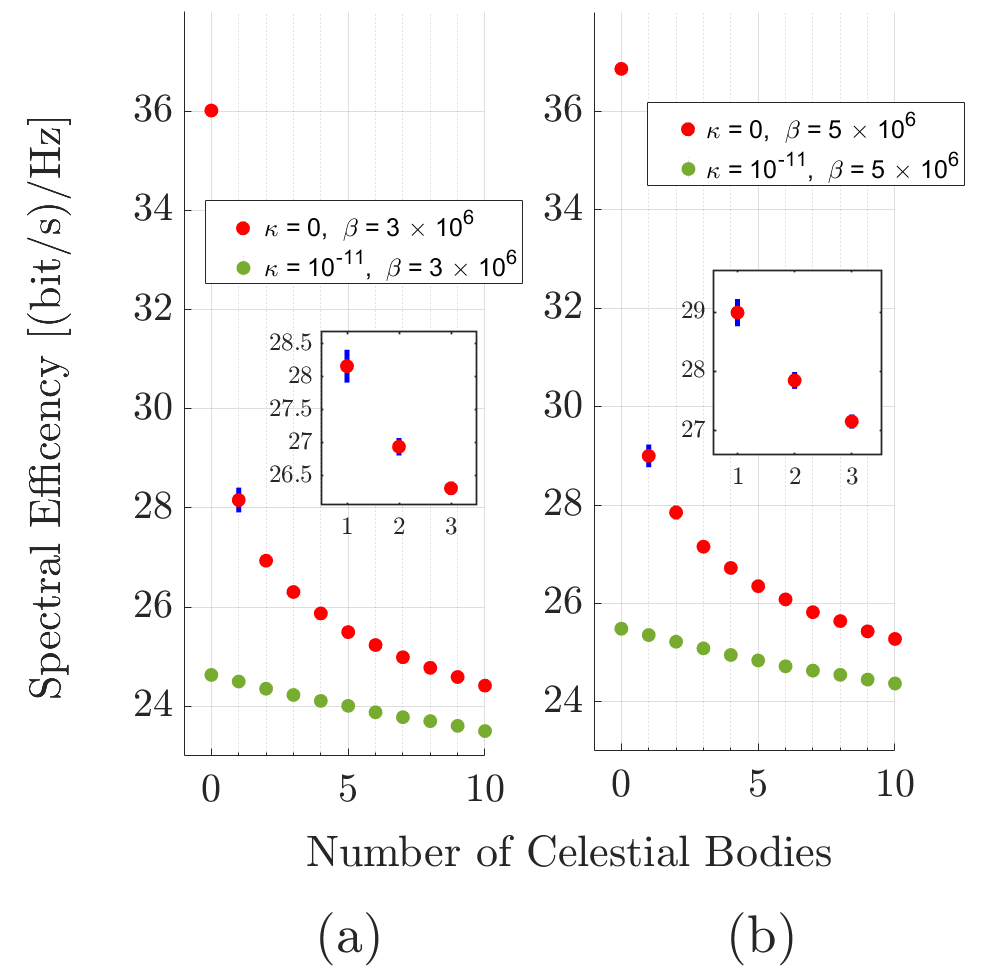}
    \caption{Suppressed celestial body source noise with low $\kappa_{1,2}$ impairments.}
    %\vspace{-5mm}
    \label{celes2}
\end{figure}

Importantly, the SE converges to a specific point for each value of $N$ in Figure \ref{celes1}. However, this is not the case for Figure \ref{celes2}, which illustrates the low hardware impairment performance under different $N$ values. Although SE has an average for $N=0$ case, the random positions and size of the celestial bodies lead to a margin of error when the hardware impairments are low. This error is indicated by the $99\%$ confidence interval (CI). The CI is determined using the Z-score and the standard deviation of the SE at the $0.99$ confidence level. 
It is observed that as $N$ increases, SE converges to an average as in a high electron density index case with high hardware impairments in Figure  \ref{celes1}. On the other hand, the SE in $\kappa=10^{-11}$ and $\beta = 3 \times 10^6$ is monotonically decreasing. 
The impact of celestial body noise sources is shown in Figure \ref{celes1} and Figure \ref{celes2}. 
\begin{table}[]
\centering
\caption{System Parameters}
\resizebox{0.6\linewidth}{!}{
\begin{tabular}{|l|l|}
\hline
Parameter & Value \\
\hline
$h_1, h_2$ $\mathrm{[m]}$ & $37786000$ \\
$f_c$ $\mathrm{[Hz]}$ & $[10\times{10^9} - 100\times{10^9}]$ \\
$\kappa_{1},\kappa_{2}$ & $[0, 10^{-11}, 10^{-10}, 10^{-8}]$ \\
$\nu_t$, $\nu_r$ & $0.782$ \\
$\theta_0$ $\mathrm{[rad]}$ & $\pi/36$ \\
$\theta_{\text{3dB}} $ $\mathrm{[rad]}$ & $0.28\pi/36$ \\
$M_1 \times M_2$ & $10400$ \\
$\beta$ & $[3\times 10^6, 5\times 10^6]$ \\
$\alpha$ & $[1 - 10]$ \\
$T$ & $100$ \\
$\omega_0$ $\mathrm{[Hz]}$ & $1$ \\
$\theta_i$ $\mathrm{[deg]}$ & $\mathbb{U}[15,30]$ \\
$P_T$ $\mathrm{[W]}$ & $[100, 500]$ \\
\hline
\end{tabular}
}
\label{reftable}
\end{table}
In the light of these findings, a comparison between the X-band and the THz band is shown in Figure \ref{xvsthz}. Observations indicate that each frequency band offers distinct advantages and disadvantages tailored to specific use cases. For example, the X-band can be advantageous due to being on a microwave window of lower black body spectral radiance than THz in the CMB. Therefore, when the CMB and celestial bodies become the dominant noise source of IP-ISL with $\kappa_1 \simeq \kappa_2 \simeq 0$, the use of X-band becomes advantageous. On the other hand, higher bands of THz are robust against plasma frequency in free space. Consequently, utilizing THz and higher bands for rather shorter communication distances (such as ISL communication within a single constellation \cite{chaudhry2020free}) offers greater advantages, thanks to reduced exposure to the deep space channels. Conversely, long-distance IP-ISL can significantly benefit from traditional microwave transmission methods. 

\section{Conclusion}

In conclusion, a reflectarray-aided IP-ISL is modeled with challenges including plasma density irregularities, hardware impairments, and hidden celestial bodies for Earth-to-Moon communication. 
\textcolor{black}{
Firstly, the thermal noise is modeled for randomly sized $N$ number of celestial bodies, and AoA fluctuations due to the dynamic plasma density channel on a $10$-GHz planar reflectarray aperture are derived. It is observed that solar plasma density impact on SE gain is heavily depending on the hardware impairments of the orbiters. Accordingly, a threshold level of $\kappa_{1,2}=10^{-8}$ is established for signal recovery without additional hardware noises. It is observed that a limited number of celestial bodies can hinder the convergence of SE to an average value, whereas SE converges as the number of celestial bodies  increases. As a result, the perfect estimation of $\delta_0$ with ideal reflectarray without $\theta_0$ limitations can fix the SE up to $38~\textrm{(bit/s)/Hz}$. On the other hand, $\theta _s \in (-\theta_0,\theta_0]$ steering range performance is constrained by the both hardware impairments and plasma density, yet presents at least $1~\textrm{(bit/s)/Hz}$ SE gain. A comparative analysis between X-band and THz frequencies is conducted based on diffraction levels across varying plasma densities for their application in IP-ISL.}     

\label{sec5}

%\bibsection*{REFERENCES}
\bibliographystyle{IEEEtran.bst}
\bibliography{reference.bib}

\begin{thebibliography}{10}
\providecommand{\url}[1]{#1}
\csname url@samestyle\endcsname
\providecommand{\newblock}{\relax}
\providecommand{\bibinfo}[2]{#2}
\providecommand{\BIBentrySTDinterwordspacing}{\spaceskip=0pt\relax}
\providecommand{\BIBentryALTinterwordstretchfactor}{4}
\providecommand{\BIBentryALTinterwordspacing}{\spaceskip=\fontdimen2\font plus
\BIBentryALTinterwordstretchfactor\fontdimen3\font minus \fontdimen4\font\relax}
\providecommand{\BIBforeignlanguage}[2]{{%
\expandafter\ifx\csname l@#1\endcsname\relax
\typeout{** WARNING: IEEEtran.bst: No hyphenation pattern has been}%
\typeout{** loaded for the language `#1'. Using the pattern for}%
\typeout{** the default language instead.}%
\else
\language=\csname l@#1\endcsname
\fi
#2}}
\providecommand{\BIBdecl}{\relax}
\BIBdecl

\bibitem{abdelsadek2022future}
M.~Y. Abdelsadek, A.~U. Chaudhry, T.~Darwish, E.~Erdogan, G.~Karabulut-Kurt, P.~G. Madoery, O.~B. Yahia, and H.~Yanikomeroglu, ``Future space networks: Toward the next giant leap for humankind,'' \emph{IEEE Transactions on Communications}, vol.~71, no.~2, pp. 949--1007, 2022.

\bibitem{balossino2022moon}
A.~Balossino and F.~Davarian, ``The {M}oon needs decent wireless coverage: Argotec and {JPL}'s relay satellites could deliver bandwidth for more than 90 missions,'' \emph{IEEE Spectrum}, vol.~59, no.~4, pp. 32--37, 2022.

\bibitem{plan2020nasa}
A.~Plan \emph{et~al.}, ``Nasa’s lunar exploration program overview,'' \emph{NASA: Washington, DC, USA}, 2020.

\bibitem{donmez2024continuous}
B.~Donmez and G.~K. Kurt, ``Continuous power beaming to lunar far side from {EMLP-2} halo orbit,'' \emph{arXiv preprint arXiv:2402.16320}, 2024.

\bibitem{israel2023lunanet}
D.~J. Israel, K.~D. Mauldin, C.~J. Roberts, J.~W. Mitchell, A.~A. Pulkkinen, D.~C. La~Vida, M.~A. Johnson, S.~D. Christe, and C.~J. Gramling, ``Lunanet: a flexible and extensible lunar exploration communications and navigation infrastructure,'' in \emph{IEEE Aerospace Conference}, pp. 1--14, 2020.

\bibitem{saeed2021point}
N.~Saeed, H.~Almorad, H.~Dahrouj, T.~Y. Al-Naffouri, J.~S. Shamma, and M.~S. Alouini, ``\textcolor{black}{Point-to-point communication in integrated satellite-aerial 6G networks: State-of-the-art and future challenges},'' \emph{IEEE Open Journal of the Communications Society}, vol.~2, pp. 1505--1525, 2021.

\bibitem{chen2019lunar}
H.~Chen, J.~Liu, L.~Long, Z.~Xu, Y.~Meng, and H.~Zhang, ``Lunar far side positioning enabled by a cubesat system deployed in an {Earth-Moon} halo orbit,'' \emph{Advances in Space Research}, vol.~64, no.~1, pp. 28--41, 2019.

\bibitem{urieta2023intelligent}
A.~Urieta-Ortega, C.~Vargas-Rosales, and R.~Villalpando-Hernandez, ``Intelligent interplanetary satellite communication network for the exploration of celestial bodies,'' \emph{Engineering Proceedings}, vol.~58, no.~1, p.~15, 2023.

\bibitem{alhilal2021roadmap}
A.~Y. Alhilal, T.~Braud, and P.~Hui, ``A roadmap toward a unified space communication architecture,'' \emph{IEEE Access}, vol.~9, pp. 99\,633--99\,650, 2021.

\bibitem{callingham2021population}
J.~Callingham, H.~Vedantham, T.~Shimwell, B.~Pope, I.~Davis, P.~Best, M.~Hardcastle, H.~R{\"o}ttgering, J.~Sabater, C.~Tasse \emph{et~al.}, ``The population of {M} dwarfs observed at low radio frequencies,'' \emph{Nature Astronomy}, vol.~5, no.~12, pp. 1233--1239, 2021.

\bibitem{ito2010asymmetric}
T.~Ito and R.~Malhotra, ``Asymmetric impacts of near-earth asteroids on the {Moon},'' \emph{Astronomy \& Astrophysics}, vol. 519, p. A63, 2010.

\bibitem{suggs2015results}
R.~Suggs and R.~M. Suggs, ``Results of lunar impact observations during {Geminid} meteor shower events,'' Tech. Rep., 2015.

\bibitem{zakharenkova2016gps}
I.~Zakharenkova, E.~Astafyeva, and I.~Cherniak, ``{GPS} and in situ swarm observations of the equatorial plasma density irregularities in the topside ionosphere,'' \emph{Earth, Planets and Space}, vol.~68, no.~1, pp. 1--11, 2016.

\bibitem{sato2018imaging}
H.~Sato, J.~S. Kim, N.~Jakowski, and I.~H{\"a}ggstr{\"o}m, ``Imaging high-latitude plasma density irregularities resulting from particle precipitation: spaceborne {L-}band {SAR} and {EISCAT} observations,'' \emph{Earth, Planets and Space}, vol.~70, pp. 1--8, 2018.

\bibitem{nishimura2021cross}
Y.~Nishimura, O.~Verkhoglyadova, Y.~Deng, and S.-R. Zhang, \emph{Cross-scale Coupling and Energy Transfer in the Magnetosphere-Ionosphere-Thermosphere System}.\hskip 1em plus 0.5em minus 0.4em\relax Elsevier, 2021.

\bibitem{lu2021effects}
L.-F. Lu, W.~Su, X.~Zhang, Z.-G. He, H.-Z. Duan, Y.-Z. Jiang, and H.-C. Yeh, ``Effects of the space plasma density oscillation on the interspacecraft laser ranging for tianqin gravitational wave observatory,'' \emph{Journal of Geophysical Research: Space Physics}, vol. 126, no.~2, p. e2020JA028579, 2021.

\bibitem{morabito2009spectral}
D.~D. Morabito, ``Spectral broadening and phase scintillation measurements using interplanetary spacecraft radio links during the peak of solar cycle 23,'' \emph{Radio Science}, vol.~44, no.~06, pp. 1--23, 2009.

\bibitem{hill2007autonomous}
K.~Hill and G.~H. Born, ``Autonomous interplanetary orbit determination using satellite-to-satellite tracking,'' \emph{Journal of guidance, control, and dynamics}, vol.~30, no.~3, pp. 679--686, 2007.

\bibitem{theoharis2022wideband}
P.~I. Theoharis, R.~Raad, F.~Tubbal, M.~U.~A. Khan, and A.~Jamalipour, ``Wideband reflectarrays for {5G/6G}: A survey,'' \emph{IEEE Open Journal of Antennas and Propagation}, vol.~3, pp. 871--901, 2022.

\bibitem{gros2021reconfigurable}
J.-B. Gros, V.~Popov, M.~A. Odit, V.~Lenets, and G.~Lerosey, ``\textcolor{black}{A reconfigurable intelligent surface at mm-Wave based on a binary phase tunable metasurface},'' \emph{IEEE Open Journal of the Communications Society}, vol.~2, pp. 1055--1064, 2021.

\bibitem{baladi2021dual}
E.~Baladi, M.~Y. Xu, N.~Faria, J.~Nicholls, and S.~V. Hum, ``Dual-band circularly polarized fully reconfigurable reflectarray antenna for satellite applications in the {Ku}-band,'' \emph{IEEE Transactions on Antennas and Propagation}, vol.~69, no.~12, pp. 8387--8396, 2021.

\bibitem{baladi2022low}
E.~Baladi, G.~S. Sethi, H.~Legay, and S.~V. Hum, ``Low-profile true-time-delay beamsteerable leaky-wave antenna for satellite applications in the {K} band,'' \emph{IEEE Transactions on Antennas and Propagation}, vol.~71, no.~1, pp. 236--249, 2022.

\bibitem{chung2023antennas}
M.-A. Chung, K.-C. Tseng, and I.-P. Meiy, ``Antennas in the internet of vehicles: Application for {X} band and {Ku} band in low-earth-orbiting satellites,'' \emph{Vehicles}, vol.~5, no.~1, pp. 55--74, 2023.

\bibitem{boulogeorgos2020much}
A.-A.~A. Boulogeorgos and A.~Alexiou, ``\textcolor{black}{How much do hardware imperfections affect the performance of reconfigurable intelligent surface-assisted systems?}'' \emph{IEEE Open Journal of the Communications Society}, vol.~1, pp. 1185--1195, 2020.

\bibitem{xu2023relay}
G.~Xu, Q.~Zhang, Z.~Song, and B.~Ai, ``Relay-assisted deep space optical communication system over coronal fading channels,'' \emph{IEEE Transactions on Aerospace and Electronic Systems}, 2023.

\bibitem{guven2023multi}
E.~G{\"u}ven, O.~B. Yahia, and G.~Karabulut-Kurt, ``Multi-state inter-satellite channel models,'' in \emph{2023 International Balkan Conference on Communications and Networking (BalkanCom)}.\hskip 1em plus 0.5em minus 0.4em\relax IEEE, 2023, pp. 1--6.

\bibitem{fang2023sensing}
X.~Fang, W.~Feng, Y.~Chen, N.~Ge, and G.~Zheng, ``Sensing-communication-computing-control closed loop for unmanned space exploration,'' \emph{arXiv preprint arXiv:2308.03658}, 2023.

\bibitem{su2019x}
T.~Su, X.~Yi, and B.~Wu, ``X/ku dual-band single-layer reflectarray antenna,'' \emph{IEEE Antennas and Wireless Propagation Letters}, vol.~18, no.~2, pp. 338--342, 2019.

\bibitem{da2018design}
M.~Da~Wu, B.~Li, Y.~Zhou, Y.~Liu, F.~Wei, X.~Lv \emph{et~al.}, ``Design and measurement of a 220 {GHz} wideband {3-D} printed dielectric reflectarray,'' \emph{IEEE Antennas and Wireless Propagation Letters}, vol.~17, no.~11, pp. 2094--2098, 2018.

\bibitem{muhleman1977electron}
D.~O. Muhleman, P.~B. Esposito, and J.~D. Anderson, ``The electron density profile of the outer corona and the interplanetary medium from {Mariner-6 and Mariner-7} time-delay measurements,'' \emph{Astrophysical Journal, Part 1, vol. 211, Feb. 1, 1977, p. 943-957.}, vol. 211, pp. 943--957, 1977.

\bibitem{durrer2020cosmic}
R.~Durrer, \emph{The Cosmic Microwave Background}.\hskip 1em plus 0.5em minus 0.4em\relax Cambridge University Press, 2020.

\bibitem{fixsen2009temperature}
D.~Fixsen, ``{The Temperature of the Cosmic Microwave Background},'' \emph{The Astrophysical Journal}, vol. 707, no.~2, p. 916, 2009.

\bibitem{ho2008solar}
C.~Ho, D.~Morabito, and R.~Woo, ``Solar corona effects on angle of arrival fluctuations for microwave telecommunication links during superior solar conjunction,'' \emph{Radio Science}, vol.~43, no.~02, pp. 1--13, 2008.

\bibitem{wheelon2001electromagnetic}
A.~D. Wheelon, \emph{Electromagnetic Scintillation: Volume 1, Geometrical Optics}.\hskip 1em plus 0.5em minus 0.4em\relax Cambridge University Press, 2001.

\bibitem{aksoy2002mixed}
A.~Aksoy and M.~Martelli, ``Mixed partial derivatives and {F}ubini's theorem,'' \emph{The College Mathematics Journal}, vol.~33, no.~2, pp. 126--130, 2002.

\bibitem{ivarsen2023measuring}
M.~F. Ivarsen and {\it et al}., ``Measuring small-scale plasma irregularities in the high-latitude {E}-and {F}-regions simultaneously,'' \emph{Scientific reports}, vol.~13, no.~1, p. 11579, 2023.

\bibitem{watson1922treatise}
G.~N. Watson, \emph{A Treatise on the Theory of Bessel Functions}.\hskip 1em plus 0.5em minus 0.4em\relax The University Press, 1922, vol.~2.

\bibitem{bjornson2013new}
E.~Bjornson, M.~Matthaiou, and M.~Debbah, ``A new look at dual-hop relaying: Performance limits with hardware impairments,'' \emph{IEEE Transactions on Communications}, vol.~61, no.~11, pp. 4512--4525, 2013.

\bibitem{nayeri2018reflectarray}
P.~Nayeri, F.~Yang, and A.~Z. Elsherbeni, \emph{Reflectarray antennas: Theory, Designs, and Applications}.\hskip 1em plus 0.5em minus 0.4em\relax John Wiley \& Sons, 2018.

\bibitem{balanis2016antenna}
C.~A. Balanis, \emph{{Antenna Theory: Analysis and Design}}.\hskip 1em plus 0.5em minus 0.4em\relax John Wiley \& Sons, 2016.

\bibitem{griffiths2006numerical}
D.~V. Griffiths and I.~M. Smith, \emph{Numerical Methods for Engineers}.\hskip 1em plus 0.5em minus 0.4em\relax Chapman and Hall/CRC, 2006.

\bibitem{chaudhry2020free}
A.~U. Chaudhry and H.~Yanikomeroglu, ``Free space optics for next-generation satellite networks,'' \emph{IEEE Consumer Electronics Magazine}, vol.~10, no.~6, pp. 21--31, 2020.

\end{thebibliography}

\end{document}